# On-orbit calibration of soft X-ray detector on Chang'E-2 satellite

Hong Xiao[a], Wenxi Peng[a*], Huanyu Wang[a], Xingzhu Cui[a], Dongya Guo[a]

[a]Institute of High Energy Physics, CAS, Beijing, China
*Corresponding author at Institute of High Energy Physics, CAS, Beijing, China

*Abstract*—X-ray spectrometer is one of the satellite payloads on Chang'E-2 satellite. The soft X-ray detector is one of the device on X-ray spectrometer which is designed to detect the major rock-forming elements within 0.5-10keV range on lunar surface. In this paper, energy linearity and energy resolution calibration is done using a weak $Fe^{55}$ source, while temperature and time effect is considered not take big error. The total uncertainty is estimated to be within 5% after correction.

## 1.Introduction

Following the precursor Chang'E-1(CE-1) satellite, a second lunar spacecraft of China, Chang'E-2 (CE-2), was launched on October 1, 2010. The CE-2 spacecraft resembled CE-1 mission with similar but the instrument performances were improved. With the onboard instruments, a huge amount of lunar scientific data were obtained successfully in CE-2 mission.

X-ray spectrometer (XRS) is one of the scientific payloads on CE-2. The XRS is composed of lunar X-ray detector (LXD) and solar X-ray monitor (SXM) which is shown in Fig.1A. The SXM is mainly used to monitor the solar X-ray flux and energy spectrum. LXD (the layout is shown in Fig.1B) is comprised of two perpendicular arrays. Each array includes two soft x-ray detectors which is used to detect the fluorescent X-rays within 1-10keV on lunar surface and eight hard x-ray detectors which is used to detect X-ray on lunar surface within energy range 25-60keV. Si-PIN X-ray sensors are used in each detector unit which equipped with bias circuits, charge sensitive preamplifiers and main amplifier circuits, while each detector unit work independently from each other. In order to monitor the performance variation of SXD, a weak $Fe^{55}$ source (~1μCi) was glued inside the collimator close to the Be window which is also shown in Fig.1B. Table 1 shows the specification of the SXDs.[1]

The scientific goal of the XRS is to obtain the abundance distribution of the major rock-forming elements (Mg, Al, Si, Ca, Ti, Fe and etc.) on lunar surface. Precise results can be got by accumulating the data in each area of the lunar surface to have enough statistics, but before this energy calibration should be done first. Energy calibration generally includes calibration of energy linearity and energy resolution. The result of on-orbit calibration using $Fe^{55}$ source will be introduced in this article.[2]

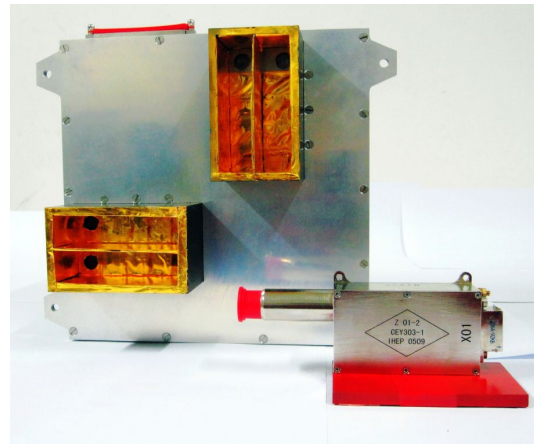

A)

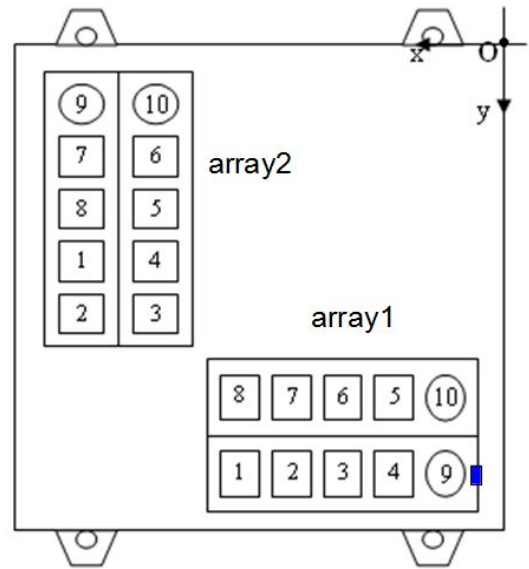

B)
Fig. 1A. X-ray spectrometer on CE-2(left: lunar x-ray detector; right: Solar X-ray monitor)
1B. X-ray spectrometer structure(circle is SXD, rectangle is HXD, blue rectangle is $Fe^{55}$ source)

Table1: the performance of SXD

| components | Soft X-ray detector(SXD) |
|---|---|
| objects | Lunar X-ray fluorescence |
| detector | Si-PIN*4 chips |
| filter | 12.5μm beryllium |
| energy range | 0.5-10 keV |
| effective area | 1cm$^2$ |
| energy resolution | 300 eV@5.9 keV |
| ADC | 10 bits |

## 2.on-orbit calibration of Energy linearity and resolution calibration

CE-2 satellite goes into circumlunar orbit from October 2nd, 2010. The data of X-ray spectrometer is transferred to the ground-based system from October 15th, 2010. About 7 months' data was obtained for this analysis, which include the message of longitude, latitude, height, quality state and etc. In this section, we select the events in good quality for the on-orbit calibration analysis from Oct. 18[th], 2010 to Oct. 20[th], 2010 which solar flare level higher than C5(solar flares higher than $5\times10^{-6}$W/m$^2$ between 1 and 8 angstroms detected by NOAA satellites).[3]

X-ray will lose energy in the sensitive medium of the detector. The detection of incident photon energy is actually the detection of photon energy lose in the detector. For the X-ray photons which deposit all the energies in the detector, the channel of electronic signal represents the incident X-ray energy. So that the relation between energy and channel can directly obtained, then channel-count spectrum can be converted into energy-count spectrum. In a word, the principle of energy linearity calibration is calibrating the correlation between different X-ray energy and the channel of electronic signal.

Mean ionization energy of semiconductor has no relationship with incident X-ray energy, so electronic signal which detector produced keep linear with incident X-ray energy.

Linear fit is used in calibration for energy-channel curve, the fitting equation is

$E = K \times ch + \Delta E$

Where Ch is channel, E is incident X-ray energy, K is slope, ΔE is intercept.[4]

The full energy peaks of Ca, Mn and Fe are fitted with gaussian to do the energy linearity and resolution calibration. The fitting result of one detector is shown in Fig.2. Four soft X-ray detectors are calibrated independently. The detailed fit result is shown in Table 2.

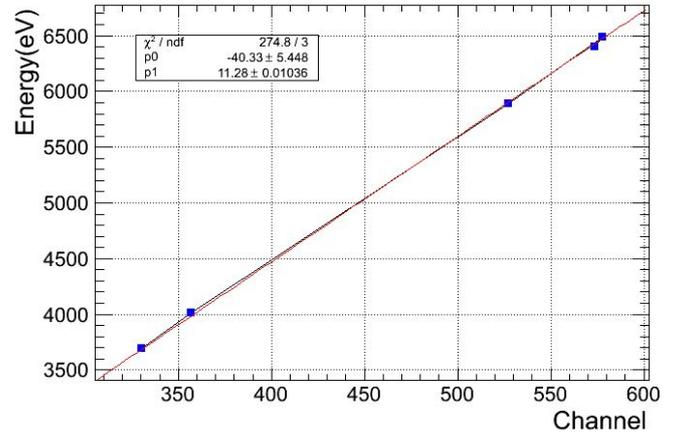

Fig. 2. The plot is the energy linearity calibration plot which shows the correlation between channel and energy.

Energy resolution of the detector is usually evaluated with FWHM of the full energy peak. The full energy spectrum of Monoenergetic X-ray in X-ray spectrometer is gaussian spectrum from theoretical estimation and calibration experiment analysis. So the energy spectrum of Si-PIN detector can be fit with gaussian distribution.

Energy resolution of X-ray spectrometer can be influenced by two factors. One is statistic fluctuation of random process in ionization, the other is caused by the noise generated by the electronic system.

The first factor is relevant to incident X-ray energy($\Delta E_1$), the relationship can be described with the function: $\Delta E_1 = (kE)^{1/2}$

The second factor has nothing to do with incident X-ray energy ($\Delta E_2$). According to the error propagation equation, FWHM can be expressed as:

$$FWHM = \sqrt{\Delta E1^2 + \Delta E2^2} = \sqrt{KE + \Delta E^2}$$

Function 3 is used in fitting energy resolution curve.

Fig. 3 is the fitting result of energy resolution calibration which shows the correlation between energy and FWHM.

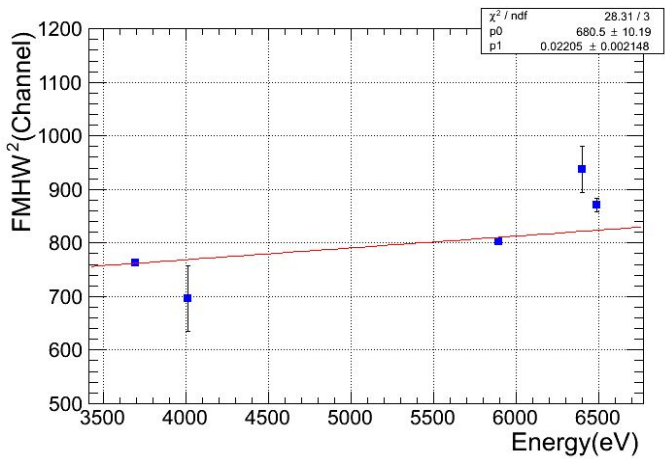

Fig. 3. The energy resolution calibration result plot.

### 3. on-orbit calibration using radioactive sources and calibration uncertainties

Because of the different environment in space, calibration result will be different. There are two factors affecting the on-orbit performance of SXD: performance changes with the orbit for short-term; space radiation damage caused by long-term degradation.

The main mechanisms of space radiation damage silicon detectors are total dose effects and displacement effects. Both effects will cause dark current increasing, the energy resolution decreasing. On circumlunar orbit, space radiation is mainly from the galactic cosmic rays, solar energetic particles and high energy electrons.

Through on-orbit calibration data with weak $Fe^{55}$ radioactive source (only affect SXD array1-9), orbit-variation and time-variation of SXD performance are analyzed in this section.

Due to there are active temperature control systems, Signal fluctuation of the Soft X-ray detector almost not affect by the outside temperature, but the electronic noise increases as the temperature increases. The energy resolution of the detector will deteriorate while temperature increases.

There are three thermal sensors (TMR130, TMR146 and TMR147) on X-ray spectrometer, which can detect the temperature of the solar X-ray monitor, lunar x-ray detector and electronic crate respectively. However, the temperature of TMR146 is the temperature of the X-ray spectrometer shell, there is an unknown temperature gradient in the vacuum environment, so it's not the exact temperature of the temperature sensitive device. Except that the parameter of the detector post-amplifier will change while time changes. So the temperature of TMR146 is just taken as reference.

Because CE-2 is a polar orbit satellite, the orbital period is about 127 minutes, there are no big change in longitude in every cycle(only about 1º), then we find there is a positive relation between the temperature of TMR146 and the latitude, about 6 hours' comparison on Oct 20th,2010 was shown in Fig. 4 for example. We select the data in a fixed latitude range, and then divide the data into several longitude bins will see the change.[5]

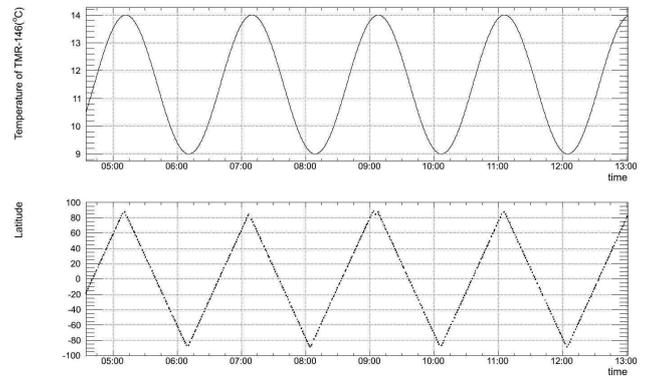

Fig. 4. The relationship between temperature and latitude on Oct 20th, 2010(top: Temperature of TMR-146 vs. time bottom: latitude vs. Time)

Powerful solar flare will affect the calibration result, so we first select the data which solar flare level lower than C5. Because temperature is related to longitude and latitude, 30º×30º small grids(longitude 12bins, latitude 6 bins) are taken to be analyzed so as to eliminate the temperature effect, which can be thought temperature almost do not change in every grid. Time effect is first checked, Mn peak in every single grid in every month is separately fitted as in Fig.5A. The relation between calibration and time is detailly shown in Fig.5B, there are six months' data is in the every individual plot, every point is the result of one month data fit, we can see no fixed regular tendency in the figure. Because of the different environment, the average energy resolution (about 28 channels) on-orbit became a little larger than ground-based result (about 26 channels) as we expected. Then latitude is fixed in range 60º to 90º to see the calibration change while the longitude changes in Fig. 5C, every point upper limit means the maximum value in 6 months in the longitude bin, while lower limit means the minimum value, and the point position is the fit result of the whole 6 months. Longitude is also fixed in range 0º to 30º to see the calibration change while the latitude changes in Fig. 5D. The detailed result of the peak position in each grid of one month is shown in Table3-4.

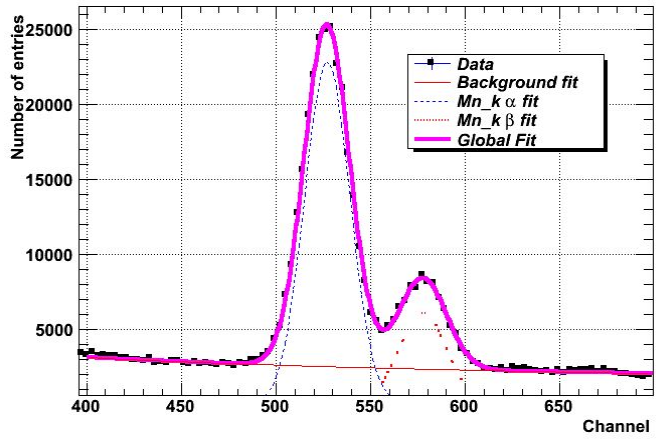
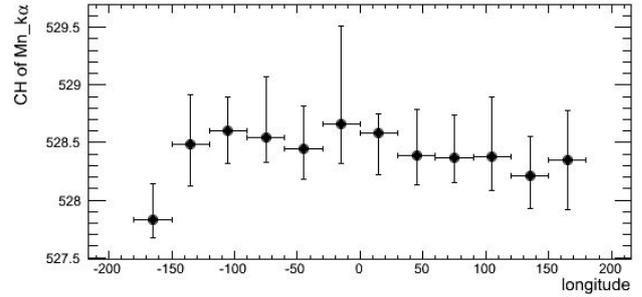
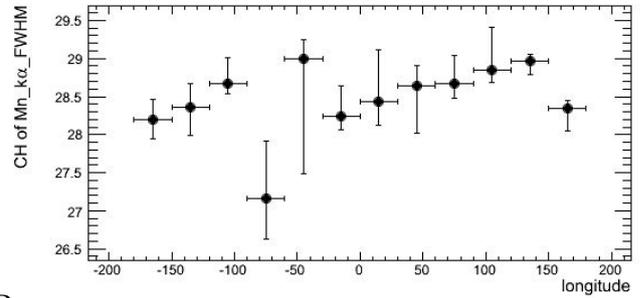

A B
C D

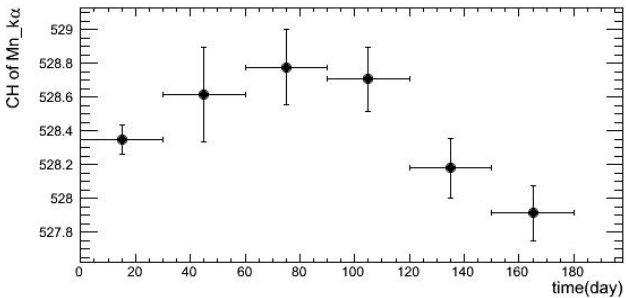
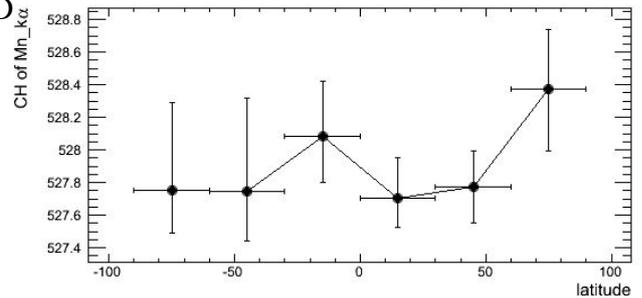
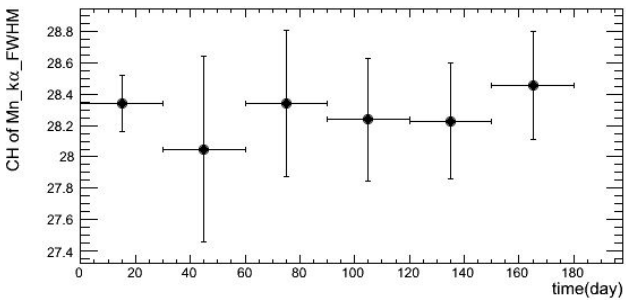
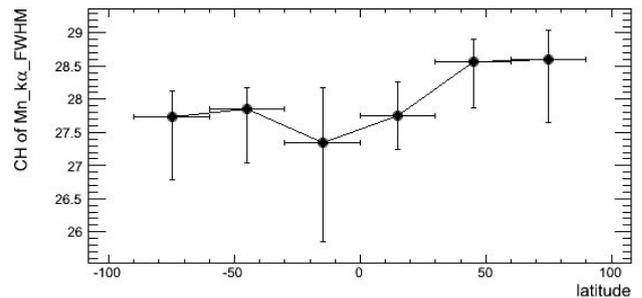

Fig. 5A. $Fe^{55}$ full energy peak fit example in one of the 30°×30° grid for one month

Fig. 5B. Calibration result change with time (top: peak position, bottom: FWHM unit:channel)

Fig. 5C. Calibration result change with longitude while latitude range within 60° to 90° (top: peak position, bottom: FWHM, unit: channel)

Fig. 5D. Calibration result change with latitude while longitude range within 0° to 30°(top: peak position, bottom: FWHM, unit:channel)

We suppose that peak position and FMHW calibration result in every month in every grid almost not affected by time, in order to get the element distribution on lunar surface, full energy peak position should be corrected with the parameters so as not to take in extra uncertainty. The efficiency detection result is supposed to take very small uncertainty. Then the systematic error of the FMHW in every grid is thought to be taken by time effect, the total error is shown in Fig. 6, which including both statistical error and systematic error, the element distribution error in the grid can be estimated by this figure. We can get the conclusion that the global calibration is within 5% during the mission.

## 4. Conclusion

In order to satisfy the scientific goal of the CE-2 project, calibrations of soft X-ray detector are well designed and successfully accomplished. On-orbit calibration are used to check the performance of each soft X-ray detectors. Energy linearity and energy resolution are calibrated using $Fe^{55}$ source. The temperature and time effect is also taken account of, while no big shift is found. The overall calibration uncertainty is estimated to be within 5%.

## 5. Acknowledgment

We thank the National Astronomical Observatories of the Chinese Academy of Sciences (NAOC) for providing the on-orbit data. We also extremely thank the reviewers for improving the paper.

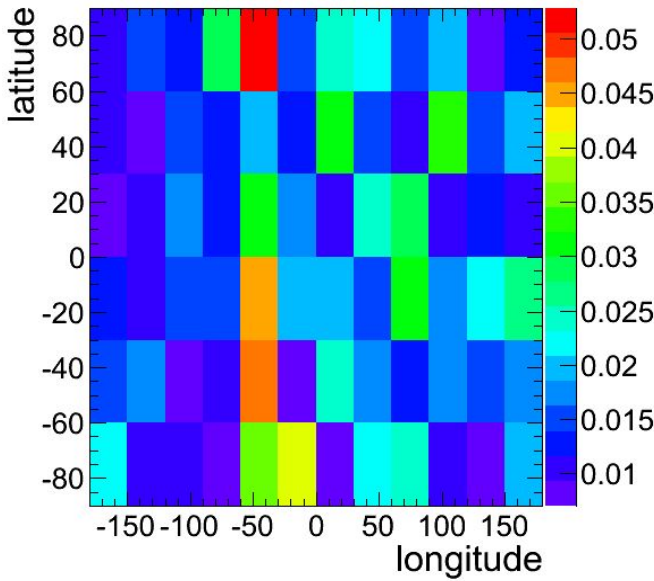

Fig. 6: energy peak and resolution vs. Latitude(left:the first month data right: the last month data)

Table 2: on-orbit calibration of energy peak and energy resolution

|  | E vs. Channel | | FWHM vs. E | |
|---|---|---|---|---|
|  | K | C | K | C |
| Array1-9 | 11.28 | -49.80 | 0.0239 | 661.89 |
| Array1-10 | 10.97 | -22.86 | 0.0386 | 637.70 |
| Array2-9 | 11.47 | 6.38 | 0.1384 | 171.82 |
| Array2-10 | 10.52 | -4.47 | -0.0411 | 993.68 |

Table 3. First month Fe-55 peak position calibration result(unit: channel)

| latitude / longitude | −90º to−60º | −60º to−30º | −30º to 0º | 0º to 30º | 30º to 60º | 60º to 90º |
|---|---|---|---|---|---|---|
| −180º to−150º | 527.732±0.086 | 527.766±0.085 | 528.008±0.087 | 527.383±0.082 | 527.495±0.085 | 527.830±0.090 |
| −150º to −120º | 527.749±0.086 | 527.879±0.084 | 527.748±0.081 | 527.616±0.082 | 527.868±0.084 | 528.486±0.087 |
| −120º to −90º | 528.184±0.078 | 528.012±0.078 | 527.967±0.085 | 528.020±0.080 | 528.430±0.088 | 528.607±0.089 |
| −90º to −60º | 528.289±0.082 | 528.207±0.087 | 528.200±0.087 | 528.080±0.090 | 528.592±0.091 | 528.539±0.095 |
| −60º to −30º | 528.047±0.103 | 527.952±0.108 | 527.665±0.108 | 527.555±0.107 | 528.384±0.114 | 528.446±0.115 |
| −30º to 0º | 527.848±0.106 | 527.946±0.106 | 527.894±0.102 | 528.134±0.101 | 528.294±0.101 | 528.660±0.099 |
| 0º to 30º | 528.164±0.093 | 527.997±0.088 | 528.418±0.087 | 528.168±0.088 | 528.286±0.087 | 528.582±0.094 |
| 30º to 60º | 527.867±0.088 | 528.220±0.086 | 527.690±0.088 | 527.742±0.088 | 528.277±0.090 | 528.391±0.091 |
| 60º to 90º | 527.775±0.083 | 527.784±0.083 | 528.081±0.084 | 527.690±0.086 | 527.745±0.084 | 528.365±0.090 |
| 90º to 120º | 527.901±0.085 | 527.718±0.089 | 527.594±0.092 | 527.652±0.089 | 527.882±0.088 | 528.381±0.086 |
| 120º to 150º | 527.752±0.085 | 527.801±0.082 | 527.628±0.084 | 528.204±0.083 | 527.731±0.082 | 528.207±0.084 |
| 150º to 180º | 527.828±0.084 | 527.912±0.082 | 527.694±0.082 | 527.971±0.085 | 527.983±0.084 | 528.350±0.086 |

Table 4. 6[th] month Fe-55 peak position calibration result(unit: channel)

| latitude \ longitude | −90º to −60º | −60º to −30º | −30º to 0º | 0º to 30º | 30º to 60º | 60º to 90º |
|---|---|---|---|---|---|---|
| −180º to −150º | 527.361±0.169 | 527.503±0.161 | 527.826±0.167 | 527.122±0.157 | 527.297±0.163 | 527.672±0.174 |
| −150º to −120º | 527.608±0.171 | 527.748±0.169 | 527.574±0.161 | 527.495±0.163 | 527.885±0.166 | 528.119±0.174 |
| −120º to −90º | 527.990±0.152 | 527.932±0.152 | 527.736±0.162 | 527.813±0.158 | 528.144±0.165 | 528.316±0.168 |
| −90º to −60º | 528.251±0.160 | 528.143±0.169 | 528.075±0.168 | 527.971±0.175 | 528.311±0.170 | 528.331±0.179 |
| −60º to −30º | 527.970±0.191 | 527.778±0.195 | 527.551±0.198 | 527.564±0.198 | 528.156±0.206 | 528.205±0.212 |
| −30º to 0º | 527.593±0.199 | 527.852±0.198 | 527.723±0.196 | 527.927±0.194 | 528.120±0.193 | 528.316±0.191 |
| 0º to 30º | 527.795±0.171 | 527.944±0.166 | 528.288±0.166 | 528.143±0.169 | 528.173±0.163 | 528.527±0.173 |
| 30º to 60º | 527.728±0.167 | 528.062±0.163 | 527.560±0.165 | 527.858±0.167 | 528.187±0.166 | 528.195±0.167 |
| 60º to 90º | 527.559±0.157 | 527.642±0.158 | 527.952±0.158 | 527.654±0.161 | 527.696±0.156 | 528.233±0.167 |
| 90º to 120º | 527.621±0.159 | 527.416±0.168 | 527.370±0.173 | 527.469±0.167 | 527.621±0.167 | 528.087±0.165 |
| 120º to 150º | 527.568±0.157 | 527.435±0.151 | 527.488±0.155 | 528.010±0.152 | 527.592±0.152 | 528.041±0.158 |
| 150º to 180º | 527.471±0.157 | 527.746±0.155 | 527.552±0.153 | 527.892±0.157 | 527.807±0.157 | 527.914±0.165 |